# An ISP Level Solution to Combat DDoS Attacks using Combined Statistical Based Approach

B. B. Gupta, Manoj Misra and R. C. Joshi

Department of Electronics and Computer Engineering
Indian Institute of Technology, Roorkee, India
*gupta.brij@gmail.com, manojfec@iitr.ernet.in and joshifcc@iitr.ernet.in*

***Abstract***: Disruption from service caused by DDoS attacks is an immense threat to Internet today. These attacks can disrupt the availability of Internet services completely, by eating either computational or communication resources through sheer volume of packets sent from distributed locations in a coordinated manner or graceful degradation of network performance by sending attack traffic at low rate. In this paper, we describe a novel framework that deals with the detection of variety of DDoS attacks by monitoring propagation of abrupt traffic changes inside ISP Domain and then characterizes flows that carry attack traffic. Two statistical metrics namely, Volume and Flow are used as parameters to detect DDoS attacks. Effectiveness of an anomaly based detection and characterization system highly depends on accuracy of threshold value settings. Inaccurate threshold values cause a large number of false positives and negatives. Therefore, in our scheme, Six-Sigma and varying tolerance factor methods are used to identify threshold values accurately and dynamically for various statistical metrics. NS-2 network simulator on Linux platform is used as simulation testbed to validate effectiveness of proposed approach. Different attack scenarios are implemented by varying total number of zombie machines and at different attack strengths. The comparison with volume-based approach clearly indicates the supremacy of our proposed system.

## 1. Introduction

Network security breaches represent a growing threat to businesses and institutions, costing them billions of dollars every year. According to statistics given by CERT [1], a mere 171 vulnerabilities were reported in 1995 that boomed to 7236 in 2007. Already, the number for the same has gone up to 4110 until the second quarter of 2008. Apart from these, a large number of vulnerabilities go unreported every year. In particular, Denial-of-Service (DoS) attacks are a major threat to the Internet. CERT defines the term "Denial of Service" as follows [2]:

*-Occupancy of limited resource or difficult to renew such as network bandwidth, data structure or memory of a system.*
*-Changeable or damage network data, for instance delete system configuration, shutdown web service.*
*-Changeable or damage physical information, for example damage of electronic, network line.*

DoS attacks are commonly characterized as events where legitimate users or organizations are deprived of certain services like web, e-mail or network connectivity that they normally expect to have. Therefore, as given by Weiler [3] they attempt:

1. To inhibit legitimate network traffic by flooding the network with useless traffic.
2. To deny access to a service by disrupting connections between two parties.
3. To block the access of a particular individual to a service.
4. To disrupt the specific system or service itself.

The main aim of such attacks is to prevent the victim either from the benefit of a particular service (in case of client being victim) or from providing its services to others (in case of server being victim).

DDoS (Distributed Denial of Service) attacks are amplified form of DoS attacks where attackers direct hundred or even more zombie machines against a single target. DDoS attacks have two phases [4]: deployment and attack phase. DDoS program must be deployed on one or more compromised hosts before attacks are possible. Thus, mitigation of DDoS attacks requires defense mechanisms for both phases. Intruder can perform DDoS attacks either as brute force attacks or as logical attacks. In brute force DDoS attacks, as shown in figure 1, legitimate looking but error data packets are sent to victim as much as possible, thus reducing legitimate user's bandwidth and preventing access to a service. Logical attacks exploit a specific feature or implementation bug of some protocol or application installed at the target machine in order to consume excess amount of its resources [4]. Series of DDoS attacks that shut down some high profile websites have demonstrated the severe consequences of these attacks [5]. A quantitative estimate of worldwide DoS attack frequency was found to be 12,000 attacks over a three-week period in 2001 [6]. As per computer crime and security survey conducted by FBI/CSI in the United States for the year 2004 [7], DoS attacks are the second most widely detected outsider attack types in computer networks immediately after virus infections. A computer crime and security survey conducted in Australia for the year 2004 [8] shows similar results.

There exist few reasons, which make DDoS attacks inevitable. The Internet is designed to keep intermediate network as simple as possible to optimize it for packets forwarding. This pushes the complexity to the end hosts and causes one unfortunate implication. If one party in two-way communication misbehaves, it can result in arbitrary damage to its peer. No one in the intermediate network will step in and stop it because Internet is not designed to police traffic. Moreover, the Internet security is highly interdependent. At the maximum we can make victim secure with firewalls etc. but still the degree of its susceptibility to DDoS attacks depends on the position of security in the rest of the global Internet [9]. The limited availability of resources acts as additional benefit for DDoS attackers. To add on, accountability is not enforced which lead to attacks comparable to reflector attacks [10] such as the Smurf attacks [11]. Thus there exists no way out to enforce global deployment of a particular security mechanism [9].

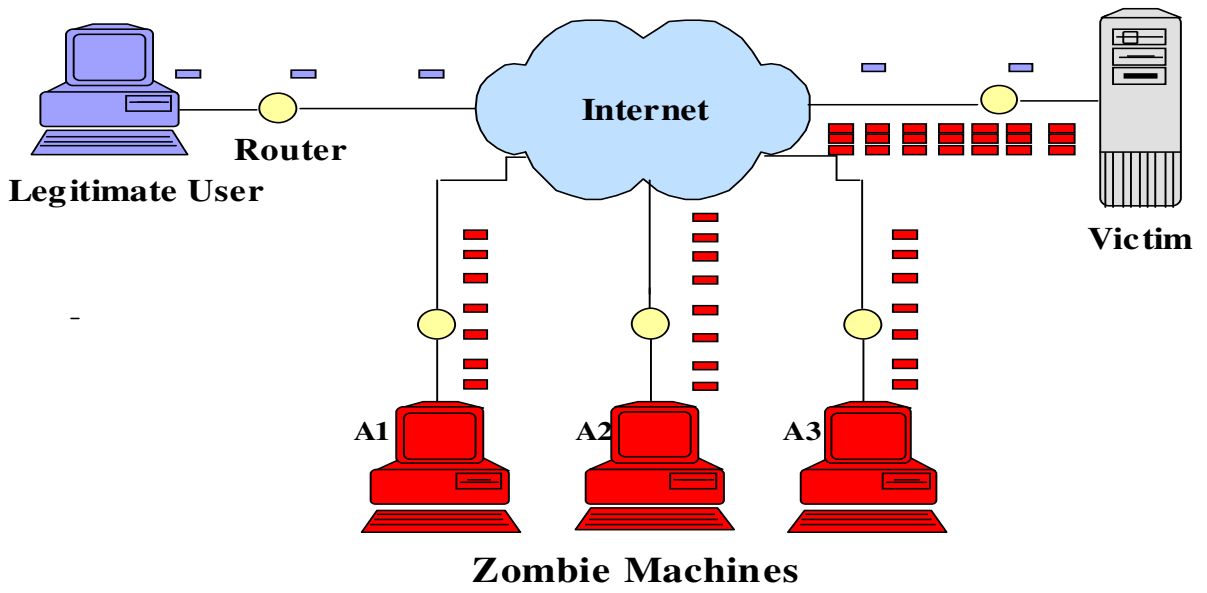

**Figure 1**. Illustration of the DDoS attack scenario

In this paper, we have proposed a novel framework that concentrates on detection and characterization of various kinds of DDoS attacks, e.g. low rate degrading, high rate disruptive and mixed rate, by monitoring the propagation of abrupt traffic changes inside ISP Domain. Two statistical metrics namely volume and flow are used as parameters to obtained normal traffic model of our system. Our proposed scheme inflicts an extremely high detection rate with low false alarm rate. Six-Sigma [12], [13] and varying tolerance factor methods are used to identify threshold values accurately and dynamically for various statistical metrics used in our scheme. Internet type topologies used for simulation are generated using Transit-Stub model of GT-ITM [14] topology generator. NS-2 [15] network simulator on Linux platform is used as simulation testbed to test our proposed scheme.

The remainder of the paper is organized as follows. Section 2 point out related work, section 3 describes our proposed approach in detail, and section 4 contains experimental design and performance analysis. Finally, Section 5 concludes the paper and outlines future work.

## 2. Related Work

This section charts out the overview on a plethora of existing DDoS defense schemes proposed in the literature.

Various reviews have been given in [4], [9], [16]-[19] on DDoS attacks, and defense methods. Molsa et al. [4] have described, what DoS attacks are, how they can be carried out in IP networks, and how one can defend against them. According to them, focus should not be to implement all possible defenses, instead, one should optimize the trade-off between security costs and acquired benefits in handling the most important risks. Mircovik et al. [9] outlined good directions for DDoS research by providing comprehensive taxonomies of attack and defense mechanisms. Moreover they critically brought forward weaknesses of various DDoS defense classes which are useful for future work in DDoS attacks field.

Peng et al. [16], have reviewed the state-of-attacks, compared the strengths and weaknesses of different defense proposals, and discussed potential countermeasures against each attack mechanism. They motivated and outlined an integrated solution to solve the problem of distributed denial of service attacks. Xiang et al. [17] have described evolution and classification of DDoS attacks. They proposed a novel concept of active defense against DDoS attacks to mitigate the infamous DDoS attacks in the Internet. Douligeris et al. [18] have presented a structural approach to the DDoS problem by developing a classification of DDoS attacks and DDoS defense mechanisms. Chen et al. [19] have proposed a characterization of DDOS defenses where reaction points were network-based and attack responses were active. And they compared different attack detection algorithms on the basis of granularity of detection used, network information monitored, specific characteristics of attack traffic, source of false positives and limitations.

Exiting DDoS defense schemes are classified into four broad categories: Prevention, Detection, Response, and Tolerance and Mitigation. Attack prevention methods try to stop all well known signature based and broadcast based DDoS attacks from being launched in the first place or edge routers, keeps all the machines over Internet up to date with patches and fix security holes. The approaches to stop IP spoofing [20], filtering malicious IP addresses based on experience [21], Remove unused services [4] and repairing security holes by patches [22] fall under this category. Attack prevention schemes are not enough to stop DDoS attacks because these are always vulnerable to novel and mixed attack types for which signatures and patches do not exist in the databases. Therefore, these are considered forensic defense methods. Attack detection aims to detect an ongoing attack and to discriminate malicious traffic from legitimate traffic. Detection can be performed using database of known signatures, by recognizing anomalies in system behaviors or using third party. Signature based approach employs a priori knowledge of attack signatures. The signatures are manually constructed by security experts analyzing previous attacks and used to match with incoming traffic to detect intrusions. SNORT [23] and Bro [24] are the two widely used signature based detection approaches. Signature based techniques are only effective in detecting traffic of known DDoS attacks whereas new attacks or even slight variations of old attacks go unnoticed. Anomaly detection [25]-[31] relies on detecting behaviors that are abnormal with respect to some normal standard. Detecting DDoS attacks involves first knowing normal behavior of our system and then to find deviations from that behavior.

Gil and Poletto [25] proposed a scheme called MULTOPS to detect denial of service attacks by monitoring the packet rate in both the up and down links. MULTOPS assumes that packet rates between two hosts are proportional during normal operation. A significant disproportion between the packet rate going to and from a host or subnet is a strong indication of a DoS attack. Blazek et al. [26] proposed batch detection to detect DoS attacks by monitoring statistical changes. Cheng et al. [27] proposed to use spectral analysis to identify DoS attack flows. Lee and Stolfo [28] used data mining techniques to discover patterns of system features that describe program and user behavior and implement a classifier that can recognize anomalies and intrusions. A mechanism called congestion triggered packet sampling and filtering is proposed by Huang et al. [29]. According to this approach, a subset of dropped packets due to congestion is selected for statistical analysis. If anomaly is indicated by the statistical results, a signal is sent to the router to filter the malicious packets. Mirkovic et al. [30] proposed D-WARD defense system that does DDoS attack detection at source, based on the idea that DDoS attacks should be stopped as close to the source as possible. Bencsath et al. [31] have given a traffic level measurement based approach, in which incoming traffic is monitored continuously and dangerous traffic intensity rises are detected. Chen et al. [32] used distributed change-point detection (DCD) architecture using change aggregation trees (CAT) to detect DDoS attack over multiple network domains. Feinstein et al. [33] focus their detection efforts on activity level and source address distribution using entropy. Anomaly based techniques can detect novel attacks; however, it may result in higher false alarms. Mechanisms that deploy third-party detection do not handle the detection process themselves, but rely on an external third-party that signals the occurrence of the attack [9]. Examples of mechanisms that use third-party detection are easily found among traceback mechanisms [34], [35].

Table 1 shows the comparison of various detection approaches i.e. pattern, anomaly and third party detection. We can see that NPSR and detection accuracy is high in pattern detection scheme compared to other. But it can be used only for known attacks detection. Effectiveness of third party detection schemes depends on detection approach used by third party. Therefore, anomaly based schemes are most efficient and effective to detect novel attacks. Because of the advantages and effectiveness over other approaches, we used anomaly based detection scheme.

The goal of the attack response is to relieve the impact of the attack on the victim while imposing minimal collateral damage to legitimate clients.

| Detection Category | Strategy Used | NPSR | Complexity | Detection Accuracy | Limitations |
|---|---|---|---|---|---|
| Pattern Detection | Store the signature of the known attacks in the databases and monitor each communication for the presence of these pattern | High | Low | High | Novel attacks detection is not possible |
| Anomaly Detection | Compare the current state of the system with normal system behavior periodically | Medium | Medium | Medium | High rate of false positive/negatives, as normal system behavior and thresholds setting is difficult |
| Third Party Detection | Rely on an third party to signals the occurrence of attack | Depend on detection approach used by third party | High | Depend on detection approach used by third party | Economic Factor, Security prone |

*Table 1.* Comparison of various Detection Approaches

The approaches to identify attack source/path or traceback [34], [35], filtering malicious traffic [36], and rate throttling malicious traffic [30], [37] fall under this category. Attack tolerance and mitigation focuses on minimizing the attack impact and tries to provide optimal level of service as per quality of its service requirement to legitimate users while service provider is under attack. The tolerance and mitigation solution includes router's queue management [38], [39], router's traffic scheduling [40], and target roaming [41].

Volume based approach (VBA) given by [31] is suitable for detection of high rate attack, but ineffective to detect low rate degrading attacks. To overcome this limitation, our scheme uses flow metric along with volume metric.

## 3. Proposed Approach

In this section, we discuss our proposed DDoS defense framework as shown in figure 2, which aims to provide the following features: (1) Detects variety of DDoS attacks while victim is being attacked, (2) identifies and tags attack flows in real time and (3) responds to identified attacks by either filtering or rate throttling according to strength of attacks.

In this paper, we will focus on successful detection of variety of DDoS attacks and characterization of malicious flows in real time. Six-Sigma method along with varying tolerance factor is used to identify threshold values correctly.

### 3.1 Detection of Attacks

After analyzing various existing DDoS defense techniques, we find that major challenges of defense against DDoS attacks are how to detect and identify the attack traffic accurately and efficiently. Detection system is part of access router or can belong to separate unit that interact with access router to detect attacks and identify attacks traffic.

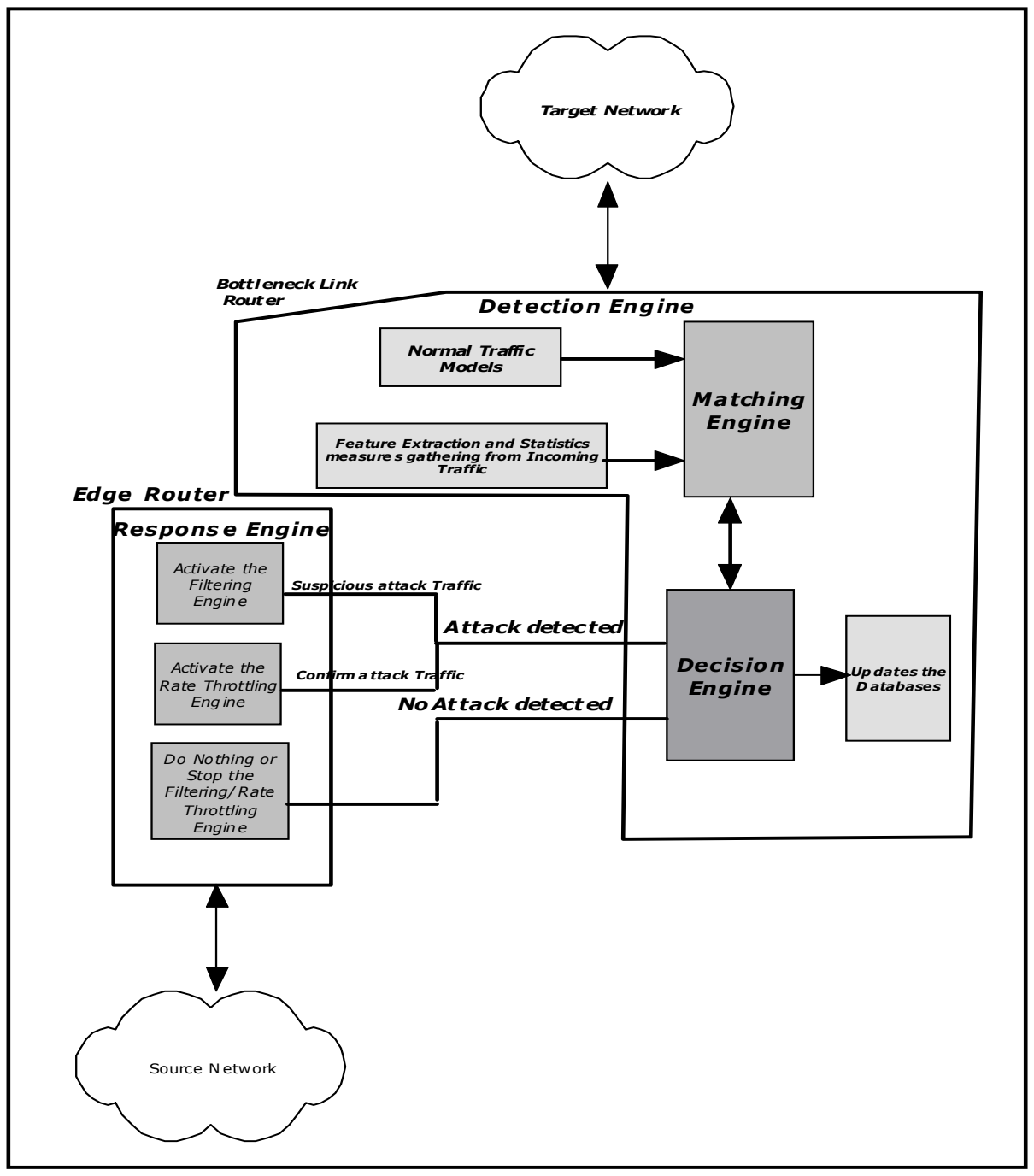

**Figure 2**. Overview of proposed DDoS Defense Framework

Detecting DDoS attacks involve first knowing normal traffic model of our system and then to find deviations from this normal traffic model. Our approach detects DDoS flooding attacks by monitoring the propagation of abrupt traffic changes inside the network. Two metrics/measures namely, Volume and Flow are used as parameter to obtained normal traffic model of our system.

Let $X_n^*(t)$ normal traffic, indicating total bytes arriving at a target machine in Δ time duration, assume that the target machine is intruded by DDoS attacks at $t_a$. Generally target may not overwhelm immediately at $t_a$. Assume attacker have attack traffic rate such that it overwhelmed target completely at $t_b$, so target is unable to provide any service to its customer. Time duration ($t_a$, $t_b$) is known as transition period of attack. A good detection approach must have detection time $t_d < t_b$, so the target may be avoided being overwhelmed completely. Let $X_{in}(t)$ be the traffic during transition period ($t_a$, $t_b$), then we can express $X_{in}(t)$ as follows:

$$X_{in}(t) = X_n^*(t) + \hat{X}(t), \quad (1)$$

In equation (1) $\hat{X}(t)$ is the component of the attack traffic. $X_{in}(t)$ - $X_n^*(t)$ using equation (1) can be used for detection purpose. Consider a random process $\{X(t), t = n\Delta, n \in N\}$, where Δ is a constant time interval, *N* is the set of positive integers, and for each *t*, X (*t*) is a random variable. Here *X* (*t*) represents the total volume in {*t* – Δ, *t*}. *X* (*t*) is calculated during time interval {*t* – Δ, *t*} as follows:

$X(t) = \sum_{i=1}^{Nf} n_i, i = 1, 2....Nf$. Here $n_i$ represent total number of bytes arrivals for a flow i in {*t* – Δ, *t*} and $Nf$ represent total number of flows. We take average of *X* (*t*) and designate that as $X_n^*(t)$ normal traffic Volume. Similarly value of flow measure is calculated and designates that as $F_n^*(t)$. Here total bytes, not packets, are used to calculate volume metric, because it provides more accuracy, as different flows can contain packets of different sizes.

To detect the attack, the value of volume metric $X_{in}(t)$ and flow metric $F_{in}(t)$ is calculated in shorter time window Δ continuously; whenever there is appreciable deviation from $X_n^*(t)$ and $F_n^*(t)$, various types of attacks are detected using algorithm 1 as given in figure 3.

Threshold values $\xi_{th}$ and $\varsigma_{th}$ are set as follows:

$$\xi_{th} = r * \sigma_V \quad (2)$$

$$\varsigma_{th} = r * \sigma_F \quad (3)$$

In equations (2) and (3), $\sigma_V$ and $\sigma_F$ represents value of standard deviation for volume and flow metrics, respectively. $r \in I$, represent value of tolerance factor. Here, *I* is a set of integers.

Effectiveness of an anomaly based detection system highly depends on accuracy of threshold value settings. Inaccurate threshold values cause a large number of false positives (legitimate traffic can be classified as attack traffic) and false negatives (attack traffic can be classified as legitimate traffic). Various simulations are performed using different value of r. Then, trade-off between detection and false positive rate provides guidelines for selecting value of r for a particular simulation environment.

***Algorithm 1: DDoS attacks Detection Algorithm***

***Input:*** $X_n^*(t)$, $F_n^*(t)$ *normal traffic Volume and Flow Metrics, respectively.* $\xi_{th}$, $\varsigma_{th}$ *threshold value for Volume and Flow Metrics, respectively.*

***Output:*** *DDoS attack alert generation.*

***Procedure:***

***01:*** *Consider a random process {* $X_{in}(t)$, $F_{in}(t)$, *t = nΔ, n∈ N}, where Δ is a constant time interval, N is the set of positive integers, and for each t,* $X_{in}(t)$ *and* $F_{in}(t)$ *are random variables. Here* $X_{in}(t)$ *represents the volume metric, and* $F_{in}(t)$ *represents the flow metric in {t – Δ, t}.*

***02:*** *If ((* $X_{in}(t) - X_n^*(t) > \xi_{th}$ *) || (* $F_{in}(t) - F_n^*(t) > \varsigma_{th}$ *))*
*DDoS attack alert is generated.*

***03:*** *Else If ((* $X_{in}(t) - X_n^*(t) < \xi_{th}$ *) && (* $F_{in}(t) - F_n^*(t) < \varsigma_{th}$ *))*
*No alert is generated.*

**Figure 3**. Algorithm for Detection of Distributed Denial of Service (DDoS) attack

So, varying tolerance factor is used to accurately and dynamically settings of threshold value. Figure 4 shows our proposed DDoS attacks detection System.

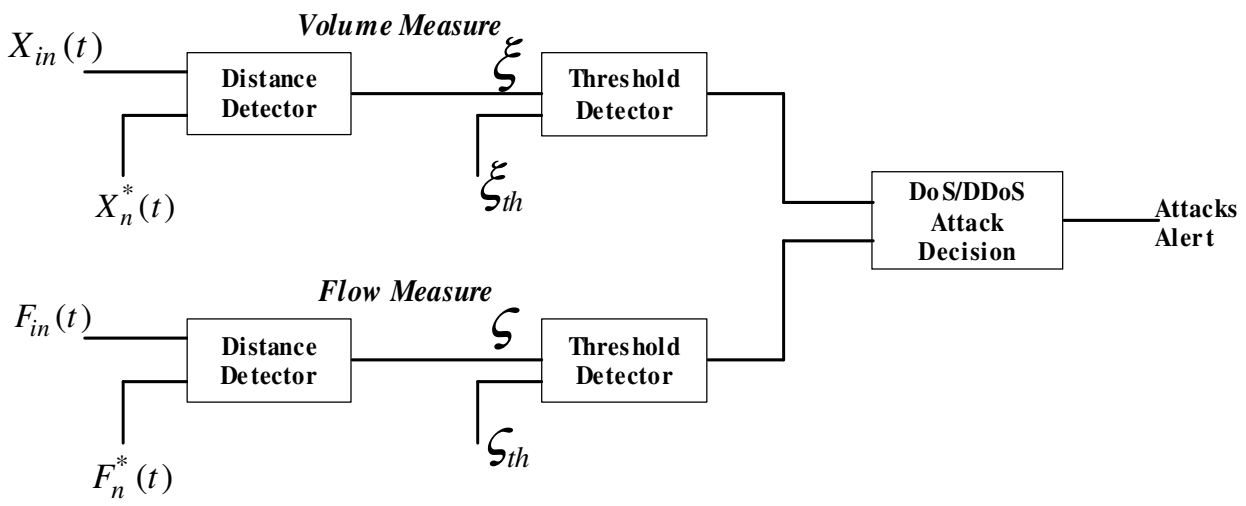

**Figure 4**. Proposed DDoS attacks detection System

### 3.2 Characterization of Malicious Flows

After detecting that DDoS attacks are occurring, next thing to do is separating traffic coming through malicious flows from legitimate traffic to respond to attacks correctly. We observed number of bytes arrival for each flow during monitoring period, and flows that crosses predefined thresholds are classify either suspicious or attack traffic flows depending on deflection from thresholds.

Let $F$ represent set of active flows. Then ($F = F_{normal} \cup F_{attack}$) AND ($F_{normal} \cap F_{attack} = \phi$), Where $F_{normal}$ represent actual normal flows and $F_{attack}$ is set of actual attack flows. Characterization algorithm outputs subsets $F_{attack}^*$, $F_{suspicious}^*$ of $F$. Here, $F_{attack}^*$ and $F^*_{suspicious}$ represent set of attack and malicious flows respectively, given by our characterization algorithm.
Ideally

$$(F^*_{attack} \cap F_{attack} = F_{attack}) AND (F^*_{attack} \cap F_{normal} = \phi)$$

and $(F^*_{suspicious} \cap F_{normal} = \phi) AND (F^*_{suspicious} \subset F)$.

Six-sigma concept is used to calculate the Upper Control Limit (UCL) and Lower Control Limit (LCL) values in order to differentiate the normal, suspicious and attack state of the total number of bytes arrival for each flow. We use the subscript ‘ss’ to represent ‘suspicious state’ and ‘as’ to represent ‘attack state’.

*3.2.1 Six-Sigma method used to identify threshold values:*

Six-Sigma, six standard deviations from the mean [12], [13], scheme is proposed by Motorola to address quality problem and business improvement. Six-Sigma means “a systematic innovative activity to statistically measure and analyze causes of defects that happen in all parts of management, and then remove those causes by identification of thresholds of the significant metrics which are measured with help of the data collected from the process”. Six-Sigma claims that focusing on reduction of variation will solve process and business problems. By using a set of statistical tools to understand the fluctuation of a process, management can begin to predict the expected outcome of that process. If the outcome is not satisfactory, associated tools can be used to further understand the elements influencing that process. Using Six-Sigma there would be approximately 3.4 or fewer failures per billion attempts. This is an extremely low rate of failure. It has been demonstrated that six sigma methodologies, integrated with rigorous statistics, can be flexible, powerful and successful without being either overly simplistic or inordinately cumbersome [42]. To find six-sigma, calculate sigma or standard deviation, multiply by 6, and add or subtract the result to the calculated mean.

Hence to achieve extremely low false positive/negative, six-sigma method is used in our attack flows characterization approach to identify the threshold values. Theoretical control limits of UCL and LCL for suspicious state are represented as:

$$UCL_{ss} = \mu + 3\sigma \quad (4)$$

$$UCL_{ss} = \mu - 3\sigma \quad (5)$$

In equations (4) and (5), $UCL_{ss}$ represents a 3 x sigma upwards deviation from the mean value of a variable. $LCLs_s$ represents a downwards 3 x sigma deviation from the mean value of a variable. For normally distributed output, 99.7% should fall between $UCL_{ss}$ and $LCL_{ss}$. Theoretical control limits of UCL and LCL for attack state are represented as:

$$UCL_{as} = \mu + 6\sigma \quad (6)$$

$$UCL_{as} = \mu - 6\sigma \quad (7)$$

In equations (6) and (7), $UCL_{as}$ represents a 6 x sigma upwards deviation from the mean value of a variable and $LCL_{as}$ represents a downwards 6 x sigma deviation from the mean value of a variable. For normally distributed output, 99.97% should fall between $UCL_{as}$ and $LCL_{as}$. Here $\mu$, $\sigma$ represent mean and standard deviation of total bytes arrival for each flow, respectively, when attack is not occurring.

Here values greater than $UCL_{as}$ or smaller than $LCL_{as}$ are considered to be under attack state. The values between $LCL_{ss}$ and $UCL_{ss}$ are considered to be under normal state. Values between $UCL_{as}$ and $UCL_{ss}$ or between $LCL_{as}$ and $LCL_{ss}$ are considered to be under suspicious state. There can still be false positives and negatives due to flash crowd. To further reduce false positive negatives, flows that are active in previous time window are omitted from list of attack flows since we assume that all attack flows start at the same time. All the packets coming through flows $F^*_{attack}$ are filtered at edge routers. Rate throttling strategy is applied to packets coming through flows $F^*_{suspicious}$. Rate of packets coming through flows $F^*_{suspicious}$ is throttled according to strength of attack. If incoming rate of attack traffic is high, packets coming through flows $F_s$ are throttle with high rate and vice versa.

# 4. Experimental Design and Performance Analysis

We tested and evaluated proposed approach with monitoring data, which is generated in our testbed to confirm its effectiveness to detect variety of Distributed Denial of service attacks.

## 4.1 Simulation Environment

The simulation is carried out using NS2 [15] network simulator. At present, the Internet can be viewed as a collection of interconnected routing domains, which are groups of nodes under a common administration that share routing information. A primary characteristic of these domains is routing locality, in which the path between any two nodes in a domain remains entirely within the domain. Thus, each routing domain in the Internet can be classified as either a stub or transit domain [43], [44]. A domain is a stub domain if the path connecting nodes u and v passes through that domain and if either u or v is located in that domain. Transit domains do not have this restriction. The purpose of transit domains is to interconnect stub domains efficiently. So, real-world Internet type topologies generated using Transit-Stub model of GT-ITM [14] topology generator is used to test our proposed scheme, where transit domains are treated as different Internet Service Provider (ISP) i.e. Autonomous System (AS). Topology contains four transit domains with each domain contain twelve transit nodes i.e. transit routers. All the four transit domains have two peer links at transit nodes with adjacent transit domains. Remaining ten transit nodes are connected to ten stub domain, one stub domain per transit node. Stub domains are used to connect transit domains with customer domains, as each stub domain contains a customer domain with ten legitimate client machines. So total of four hundred legitimate client machines are used to generate background traffic. Total zombie machines range between 10 and 100 to generate attack traffic. Transit domain four contains the server machine to be attacked by zombie machines. A short scale simulation topology is shown in figure 5.

The legitimate clients are TCP agents that request files of size 1 Mbps each with request inter-arrival times drawn from a Poisson distribution. The attackers are modeled by UDP agents.

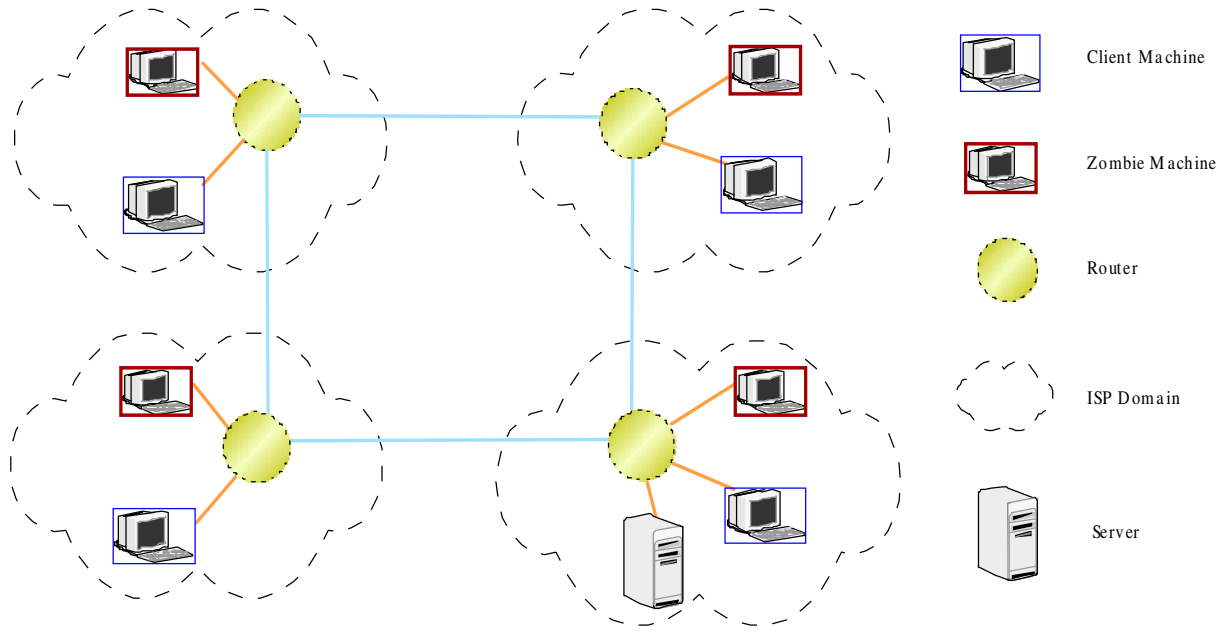

**Figure 5**. A short scale simulation topology

A UDP connection is used instead of a TCP one because in a practical attack flow, the attacker would normally never follow the basic rules of TCP, i.e. waiting for ACK packets before the next window of outstanding packets can be sent, etc. The attack traffic rate varies from 0.1 to 3.5 Mbps per attack daemon. The size of monitoring window affects the number of attack alert raised. In our experiments, the monitoring time window was set 200 ms, as the typical domestic Internet RTT is around 100 ms and the average global Internet RTT is 140 ms [45]. Using this value of monitoring window, total numbers of false positive alarms are minimum. False positive alarm number increases steadily with increasing monitoring window size. The simulations are repeated and different attack scenarios are compared by varying total number of zombie machines and at different attack strengths.

Figure 6 shows temporal variation of volume metric when (a) system is in normal condition, (b) under low rate DDoS attack and (c) under high rate DDoS attack.

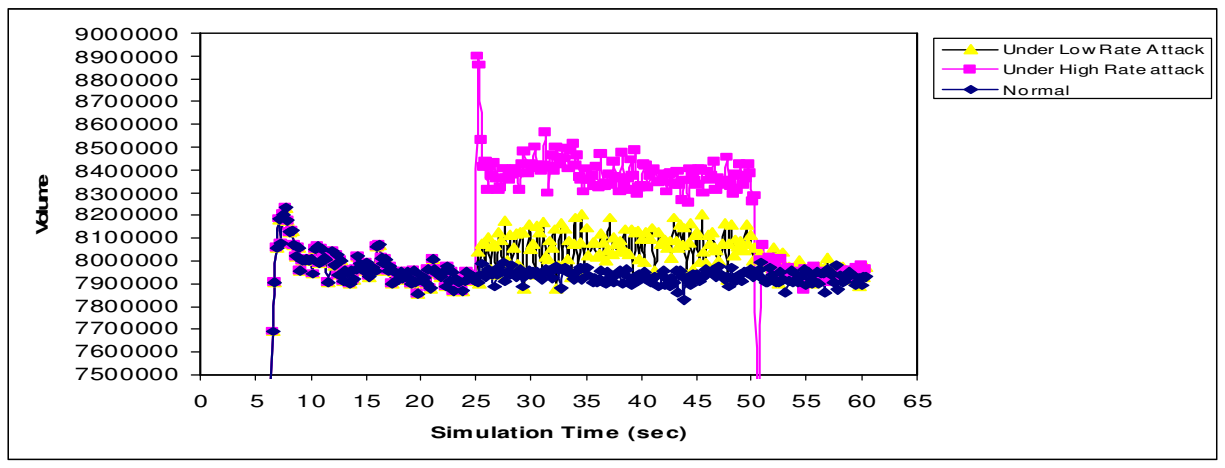

**Figure 6**. Temporal variation of volume metric when system is in normal condition, under low rate DDoS attack, and under high rate DDoS attack

DDoS attacks start at 25th second and end at 50th second. 400 client machines are used to send TCP traffic. High rate attack is performed using 100 zombie machines with mean rate 3Mbps per attacker. To perform low rate attack 100 zombie machines are used with mean rate 0.1Mbps per attacker. As shown in figure, it is clear that low rate attacks are nearly undetectable when using only volume as statistical measure.

For detection of low rate DDoS attack correctly with low false positive rate, flow metric is also considered along with volume metric. Figure 7 shows temporal variation of flow metric when (a) system is in normal condition, (b) under DDoS attack using 25, 50, 75 and 100 zombie machines.

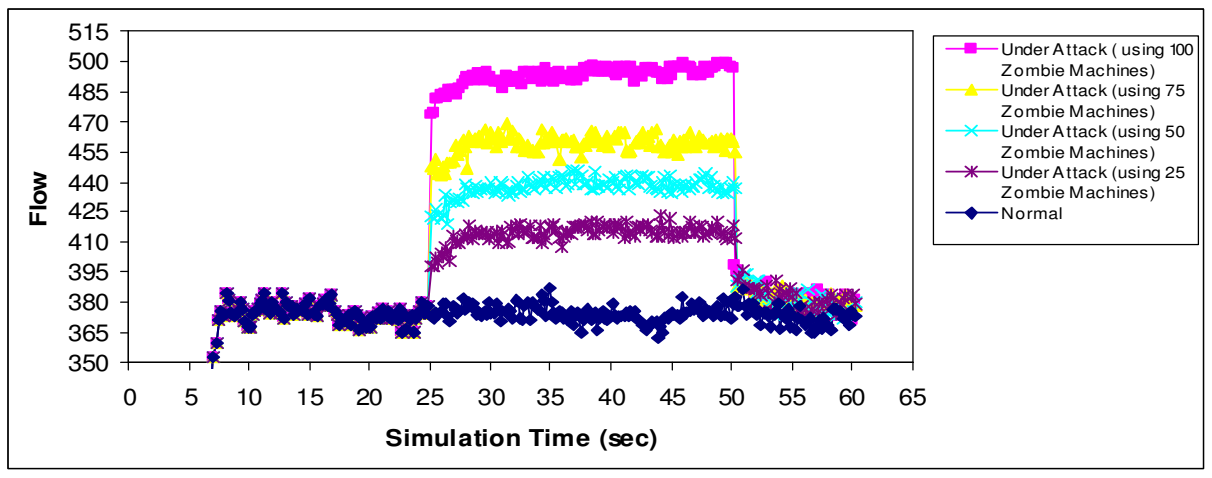

**Figure 7**. Temporal variation of Flow metric when system is in normal condition, and under DDoS attack using 25, 50, 75 and 100 zombie machines

It is clear from the figure 6 and figure 7, that low rate DDoS attacks perform using large number of zombie machines are also easily detected using both flow and volume metrics simultaneously.

### 4.2 Performance Evaluation Metrics

We have used three metrics to evaluate performance of our proposed DDoS detection approach, namely, detection rate ($R_d$), false positive alarm rate ($R_{fp}$), and receiver operating characteristic (ROC). The detection rate ($R_d$) is the measure of percentage of attacks detected among all actual attacks performed. The detection rate ($R_d$) is defined as follows:

$$R_d = d/n \qquad (8)$$

Where d is the number of DDoS detected attacks, and n is the total number of actual attacks generated during the simulation. The false positive alarm rate ($R_{fp}$) is the measure of percentage of false positives among all normal traffic event defined as follows:

$$R_{fp} = f/m \qquad (9)$$

Where f is the number of false positive alarm raised by attack detection mechanism, and n is the total number of normal traffic flow events during the simulation. The ROC curve is used to evaluate tradeoff between detection rate and false positive rate.

### 4.3 Simulation Results and Discussion

Figure 8 illustrates the variation of the detection and false positive rate with respect to different value of detection tolerance factor r, when DDoS is perform using different packet size. Detection rate is close to 100% with r <=6 and False positive alarm rate is <=2.9% with r >=6.

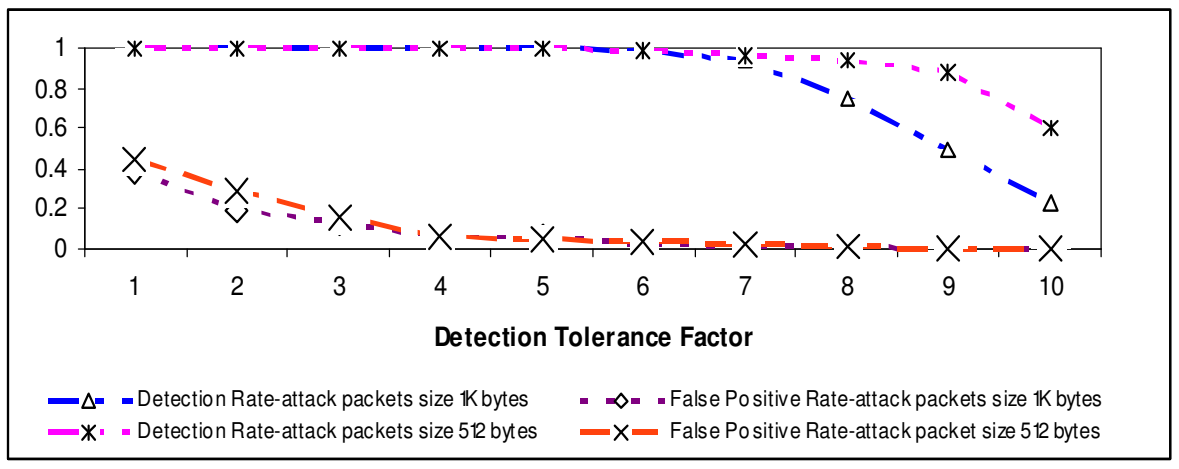

**Figure 8**. Effect of detection tolerance factor on the detection and false positive rate

Above result demonstrates that detection rate is very high with low false positive rate when r=6. The ROC curve in figure 9 explains the tradeoff between the detection rate and the false positive rate when DDoS is performing using different packet size. Our detection scheme achieves a detection rate as high as 98.4% with 2.9% false positive rate. At detection rate 94.4%, false positive rate is very low 1.8%. So value of r is taken 6 in our approach. Value of r varies according to different simulation environments and correct value can be selected by drawing tradeoff between detection and false positive rate.

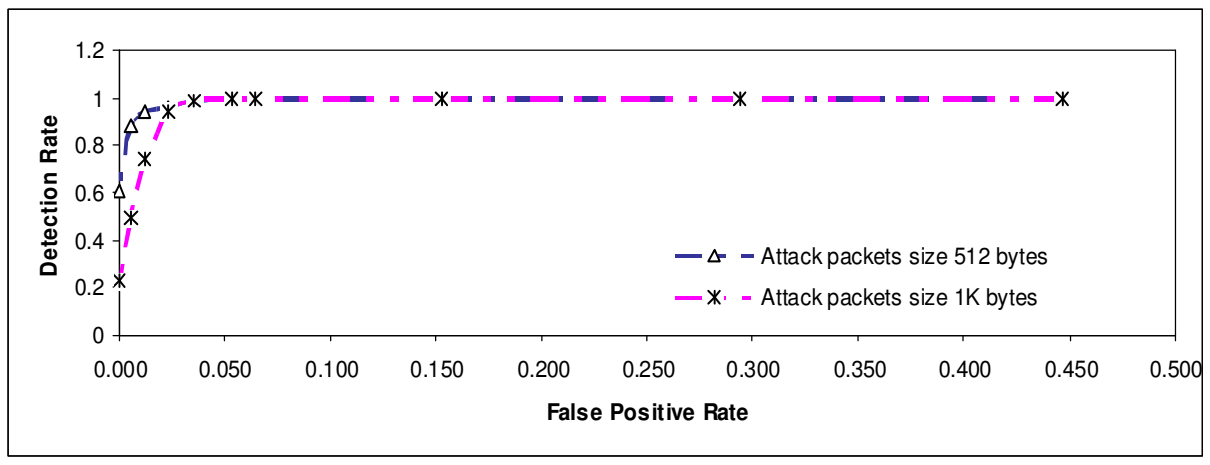

**Figure 9**. ROC curve showing the tradeoff between the detection rate and false positive rate of DDoS attacks

### 4.4 Comparison with VBA

Comparison of detection performance of our proposed approach with VBA (Volume Based Approach) [31] DDoS attack detection system is reported below. VBA is implemented in our testbed. Following different DDoS attack scenarios are taken for comparison:

*4.4.1 Attack with high rate is performed by varying number of zombie machines*

To completely disrupt services provided by server machine or to high degradation of performance of sever machine, attack with high rate (300 Mbps) is performed by attacker. False positive rate is comparable in both the cases. Figure 10 have shown the variation of detection rate of VBA and our detection system when attack is performed by varying total number of zombie machines.

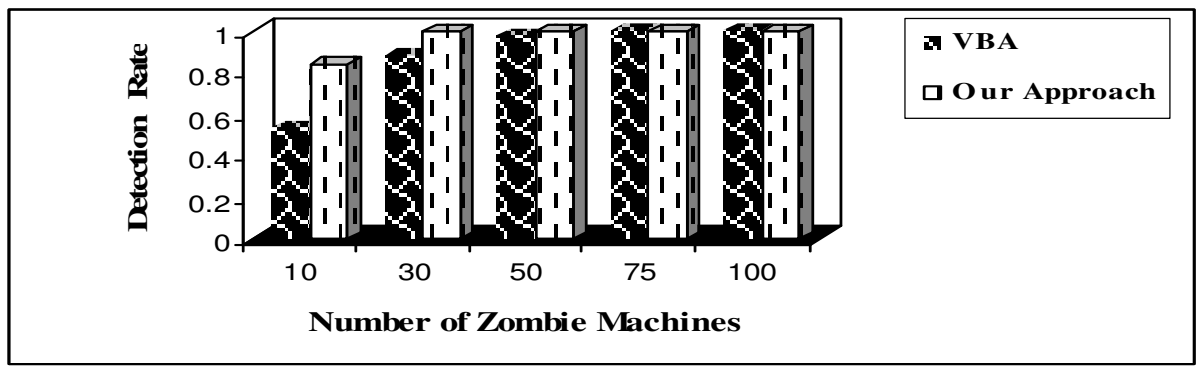

**Figure 10**. Variation of detection rate of VBA and our detection system when attack with high rate is performed by varying number of zombie machines

It is demonstrated by the figure that in this case detection results are comparable with more zombie machines, but when total number of zombie machines are less, our approach provide better detection rate compare to volume based approach.

*4.4.2 Attack with low rate is performed by varying total number of zombie machines*

To low degradation of performance of server machine, attack with low rate (10 Mbps) is performed by attacker. Figure 11 have shown the variation of detection rate of VBA and our detection system when attack is performed by varying number of zombie machines. It is demonstrated by the figure that our detection system's performance is far ahead and better than VBA.

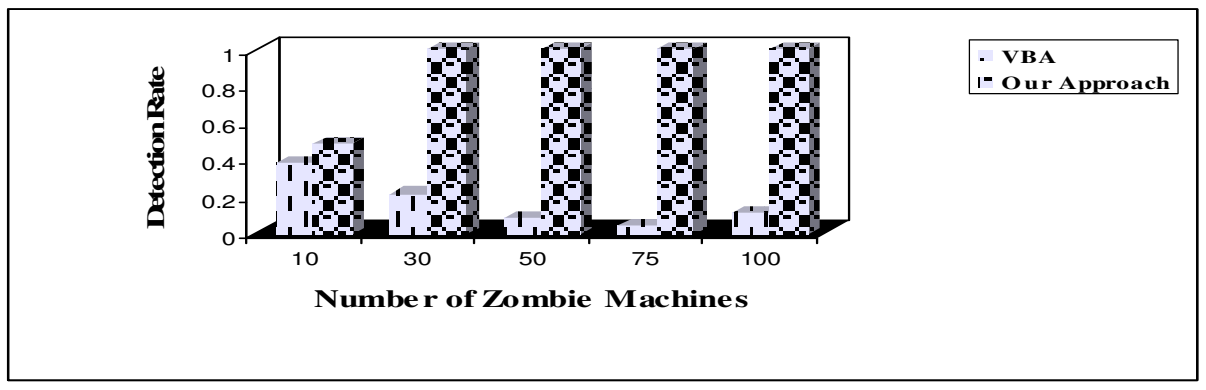

**Figure 11**. Variation of detection rate of VBA and our detection system when attack with low rate is performed by varying number of zombie machines

This is mainly due to the fact that in case of low rate degrading attacks the total arrived attack traffic does not exceed even normal fluctuation. But as we have considered total arrival flows too with arrival traffic, low rate degrading attacks are easily detected by our approach.

*4.4.3 Attack with varying attack rate is performed using fixed number of zombie machines*

Here we have considered varying attack rate using 100 zombie machines to degrade performance of server machine.

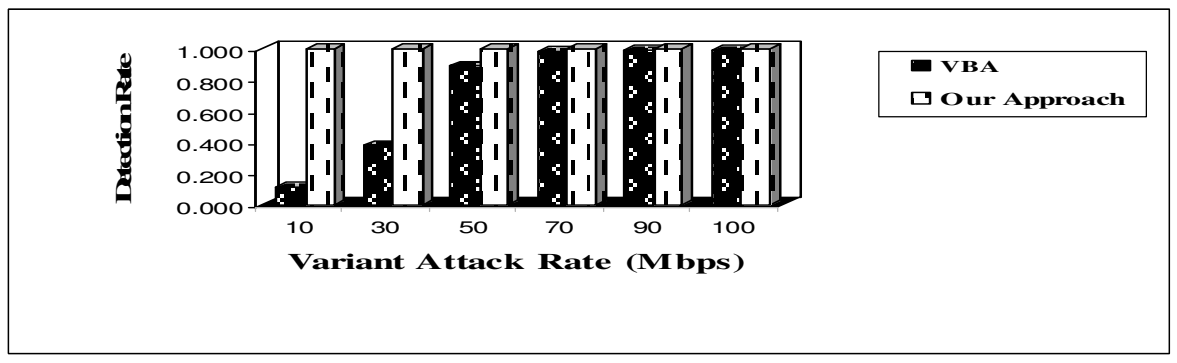

**Figure 12**. Variation of detection rate of VBA and our detection system when attack with varying attack rate is performed using hundred zombie machines

It is demonstrated by the figure 12 that in this our detection system's performance if far ahead and better than VBA when attack strength is low. This is mainly due to the fact that the total arrived attack traffic does not exceed even normal fluctuation.

*4.4.4. Attack is performed when variation in both client and attack load*

Here we have considered the case when both client and attack load vary i.e. attack and client load is low, moderate and high. As shown in figure 13, it is clear that our detection system's performance is far ahead and better than VBA.

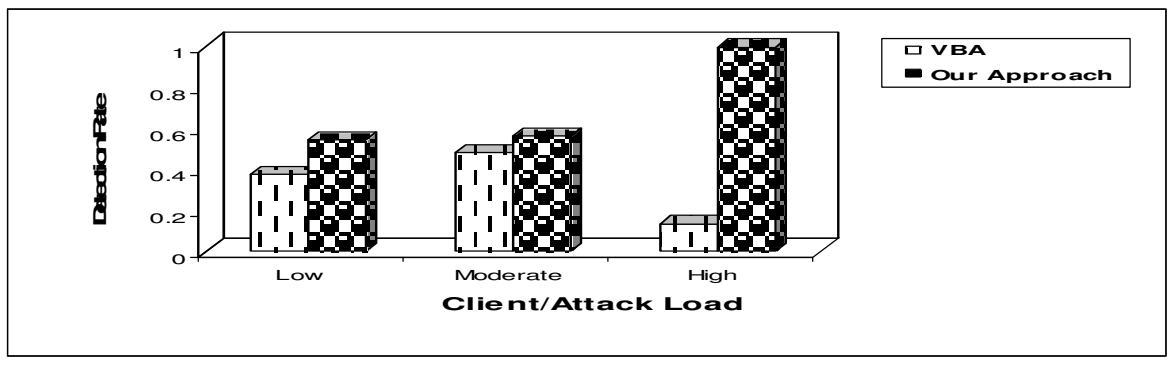

**Figure 13**. Variation of detection rate of VBA and our detection system when attack with low rate is performed by varying number of zombie machines

## 5. Conclusion and Future Work

In this paper, we have proposed a novel framework that deals with the detection of variety of DDoS attacks i.e. high rate, low rate, mixed rate etc.; by monitoring the propagation of abrupt traffic changes inside ISP Domain and characterizes flows that carry attack traffic. Then its effectiveness is verified through intensive experiments on our testbed. We have shown by simulation results that, our novel framework can effectively detect and characterize various kinds of DDoS attacks with extremely high detection rate and with low false alarms rate. Although simulation results are promising, but in future work we plan to validate our approach, with real datasets. Investigation of an accurate strategy for response to identified attacks is also a future research issue.

## Author Biographies

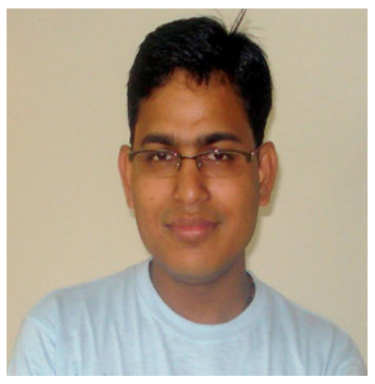

**B. B. Gupta** received the bachelor’s degree in Information Technology in 2005 from Rajasthan University, India. He is currently a PhD student in the Department of Electronics and Computer Engineering at Indian Institute of Technology, Roorkee, India. His research interests include defense mechanisms for thwarting Denial of Service attacks, Network security, Cryptography, Data mining and Data structure and Algorithms.

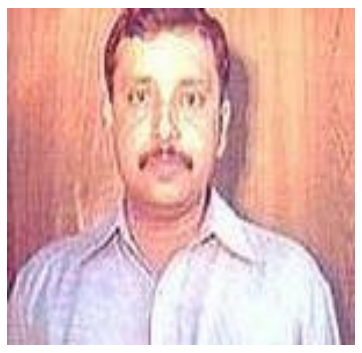

**Manoj Misra** received the bachelor’s degree in Electrical Engineering in 1983 from HBTI Kanpur, India. He received his master’s and PhD degree in Computer Engineering in 1986 and 1997 from University of Roorkee, India and Newcastle upon Tyne, UK, respectively. He is currently a Professor at Indian Institute of Technology Roorkee. He has guided several PhD theses, M.E./M.Tech. Dissertations and completed various projects. His areas of interest include Mobile computing, Distributed computing and Performance Evaluation.

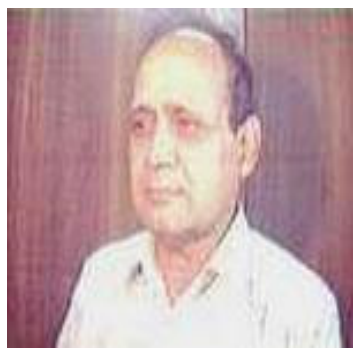

**R. C. Joshi** received the bachelor’s degree in Electrical Engineering from Allahabad University, India in 1967. He received his master’s and PhD degree in Electronics and Computer Engineering from University of Roorkee, India in 1970 and 1980, respectively. He is currently a Professor at Indian Institute of Technology Roorkee, India. He has a vast teaching experience exceeding 38 years at graduate and postgraduate levels at IIT Roorkee. He has guided over 150 M.Tech and 25 PhD dissertations. He has published over 100 research papers at national and international journals and presented many in Europe, USA and Australia. He has been awarded Gold Medal by Institute of Engineers for best paper. He has chaired many national and international conferences and workshops. Presently, he is actively involved in research in the field of Database management system, Data mining, Bioinformatics, Information security, Reconfigurable systems and Mobile computing.